\documentclass[aps,pra,twocolumn,superscriptaddress,floatfix]{revtex4-2}

\usepackage{amssymb,amsmath}
\usepackage{graphicx}
\usepackage{dcolumn}
\usepackage{bm}
\usepackage[dvipsnames]{xcolor}
\usepackage{hyperref}
\hypersetup{colorlinks=true,urlcolor=[RGB]{39 110 151},citecolor=black,linkcolor=black,pdfborder={0 0 0}}
\usepackage{environ}
\usepackage{braket}
\usepackage{siunitx}
\DeclareSIUnit\gauss{G}

\newcommand{\hd}{0.31(1)}
\newcommand{\hmax}{0.53(1)}
\newcommand{\DeltaMain}{14(2)}
\newcommand{\DeltaFigOffset}{15(1)}

\newcommand{\uzero}{-0.2(3)}

\newcommand{\usix}{6.6(1)}

\newcommand{\uten}{10.0(1)}

\newcommand{\Tzero}{0.30(5)}

\makeatletter
\newwrite\remember@figures
\AtBeginDocument{\InputIfFileExists{\jobname.dft}{}{}\immediate\openout\remember@figures=\jobname.dft}
\AtEndDocument{\immediate\closeout\remember@figures}
\NewEnviron{dfigure}[1]{%
    \immediate\write\remember@figures{%
        \noexpand\rememberfigure{#1}{\unexpanded\expandafter{\BODY}}%
    }%
}
\NewEnviron{dfigure*}[1]{%
    \immediate\write\remember@figures{%
        \noexpand\rememberfiguretc{#1}{\unexpanded\expandafter{\BODY}}%
    }%
}
\newcommand{\placefigure}[2][tp]{%
    \csname remembered@figure@#2\endcsname{#1}%
}
\newcommand{\rememberfigure}[2]{%
    \global\@namedef{remembered@figure@#1}##1{%
        \begin{figure}[##1]#2\end{figure}%
    }%
}
\newcommand{\rememberfiguretc}[2]{%
    \global\@namedef{remembered@figure@#1}##1{%
        \begin{figure*}[##1]#2\end{figure*}%
    }%
}
\makeatother

\begin{document}

\title{Ferrimagnetism of ultracold fermions in a multi-band Hubbard system}

\newcommand{\harvard}{Department of Physics, Harvard University, Cambridge, MA 02138, USA}
\author{Martin Lebrat}
\thanks{These authors contributed equally to this work}
\author{Anant Kale}
\thanks{These authors contributed equally to this work}
\author{Lev Haldar Kendrick}
\author{Muqing Xu}
\author{Youqi Gang}
\author{Alexander Nikolaenko}
\author{Pietro M. Bonetti}
\author{Subir Sachdev}
\author{Markus Greiner}
\thanks{\href{mailto:mgreiner@g.harvard.edu}{\color{black}{mgreiner@g.harvard.edu}}}
\affiliation{\harvard}

\date{\today}

\begin{abstract}

Strongly correlated materials feature multiple electronic orbitals which are crucial to accurately understand their many-body properties, from cuprate materials to twisted bilayer graphene.
In such multi-band models, quantum interference can lead to flat energy bands whose large degeneracy gives rise to itinerant magnetic phases, even with weak interactions.
Here, we report on signatures of a ferrimagnetic state realized in a Lieb lattice with ultracold fermions, characterized by antialigned magnetic moments with antiferromagnetic correlations, and concomitant with a finite spin polarization.
We demonstrate their robustness when increasing repulsive interactions from the weakly-interacting to the Heisenberg regime, and show their emergence when continuously tuning the lattice unit cell from a square to a Lieb geometry.
Our flexible approach to realize multi-orbital models paves the way towards exploring exotic phases such as quantum spin liquids in kagome lattices and heavy fermion behavior in Kondo models.

\end{abstract}

\maketitle

\section*{Introduction}

Multi-orbital effects in quantum materials are often critical to fully understand their many-body properties, but can be challenging to predict theoretically and numerically. Cuprate materials, for example, are widely thought to be amenable to a two-dimensional Hubbard model describing the tunneling and interactions of holes within copper-oxygen planes \cite{anderson_resonating_1987,zhang_effective_1988}. Yet, a detailed microscopic account of the hybridization of oxygen and copper orbitals, which involves multiple energy bands \cite{emery_theory_1987,varma_charge_1987}, may be key to fully explain the emergence of a $d$-wave superconducting state \cite{rybicki_perspective_2016,simons_absence_2020,kowalski_oxygen_2021,jiang_density_2023}.

The Lieb lattice represents a paradigmatic example of a multi-band Hubbard system.
It can be viewed as a simplified description of the cuprate lattice, with a unit cell containing two $p$-sites and one $d$-site that features a uniform nearest-neighbor tunneling $t$, on-site interaction $U$ and potential energy (Fig.~\ref{fig:flat_band}A).
Because the Lieb lattice is bipartite and its two sublattices have an unequal number of sites, its ground state for spin-$1/2$ particles at half-filling has a non-zero total spin $S = 1/2$ times the number of unit cells for any interaction $U > 0$, as rigorously shown by Lieb \cite{lieb_two_1989}. 
This macroscopic spin is however associated with long-range antiferromagnetic correlations \cite{shen_ferrimagnetic_1994} with opposite magnetic moments on the two sublattices, making the Lieb lattice a prime host for \emph{ferrimagnetism}.

The existence of this magnetic state over a broad range of interactions is intimately connected to the presence of a flat band (Fig.~\ref{fig:flat_band}B), whose degeneracy is lifted by infinitesimal interactions \cite{auerbach_interacting_2012,moriya_itenerant_magnetism} and whose properties are deeply influenced by quantum geometry and topology
\cite{huber_bose_2010,goldman_topological_2011,tang_high-temperature_2011,neupert_fractional_2011,sun_nearly_2011,peotta_superfluidity_2015,peotta_quantum_2023}.
Interest in flat-band systems has been spurred by the recent discovery of kagome metals \cite{kang_dirac_2020,han_evidence_2021} and exotic phases in twisted bilayer graphene, in which correlations are strongly enhanced by flat minibands \cite{bistritzer_moire_2011,andrei_graphene_2020}. 
Dispersionless bands have been further engineered in artificial electronic lattices \cite{drost_topological_2017,slot_experimental_2017} and with bosonic particles in a variety of experimental platforms \cite{leykam_artificial_2018}, including photonic systems \cite{guzman-silva_experimental_2014,vicencio_observation_2015,mukherjee_observation_2015}, exciton-polariton gases \cite{whittaker_exciton_2018} and superconducting circuits \cite{martinez_flat-band_2023}.

Ultracold atoms in optical lattices and tweezers provide a clean experimental realization of flat-band models \cite{jo_ultracold_2012,taie_coherent_2015,taie_spatial_2020,leung_interaction-enhanced_2020,li_aharonov-bohm_2022,wei_observation_2023,chen_interaction-driven_2024}.
However, achieving the strongly correlated regime with fermions required to uncover their emergent magnetic properties \cite{scalettar_antiferromagnetic_1991,iglovikov_superconducting_2014,costa_ferromagnetism_2016,kumar_temperature_2017,oliveira-lima_dynamical_2020} has been challenging so far.
In this work, we perform the quantum simulation of a multi-band Hubbard model in an optical Lieb lattice, using fermionic lithium-6 atoms with interactions that are widely tunable with a Fano-Feshbach resonance.
In contrast to bosonic platforms, the Pauli exclusion principle associated with fermions allows us to populate and characterize a non-ground flat band at equilibrium and across half-filling. Through site-resolved measurements of spin, we observe unequal and opposite magnetizations of the Lieb sublattices when subjecting the system to a global Zeeman field \cite{supplementary}. This differential magnetic susceptibility, together with antiferromagnetic spin correlations, represent finite-temperature signatures of a ferrimagnetic ground state at zero field that persist when increasing interactions from the metallic to the insulating regime.

\placefigure{f1}

\section*{Flat-band lattice}

In the absence of interactions, the large state degeneracy of the Lieb lattice can be interpreted as the presence of spatially bounded eigenstates over each single square plaquette \cite{sutherland_localization_1986}. 
As shown in Fig.~\ref{fig:flat_band}B, these degenerate eigenstates are localized on the $p$-sublattice owing to destructive quantum interference between tunneling events on the $d$-sublattice. 
As a result, the flat band constructed from these eigenstates is also expected to be localized on the $p$-sublattice,
effectively forming a dark state of the tight-binding Hamiltonian written as a three-level system in momentum-space \cite{supplementary,taie_spatial_2020}.

Experimentally, we realize a Lieb lattice by decorating every fourth site of an optical square lattice using a $5\times5$ array of repulsive potentials projected through a high-resolution microscope objective \cite{supplementary}. The Lieb lattice is tunnel-coupled to a non-decorated square lattice acting as a reservoir (Fig.~\ref{fig:flat_band}C), which allows us to both set and measure the chemical potential and temperature using the density and spin correlations of the reservoir \cite{supplementary}. 

We make use of the fermionic nature of the atomic cloud and our ability to change its chemical potential to probe the peculiar band structure of the engineered Lieb lattice. 
We first work in the non-interacting limit $U/t=\uzero$ by tuning out the interactions between the atoms via a Fano-Feshbach resonance.
We measure the average atomic density in the Lieb lattice for different total atom numbers loaded into the trap, and relate it to the global chemical potential by inverting the known equation of state in the square lattice (Fig.~\ref{fig:flat_band}D, \cite{supplementary}). Differentiating the local density versus chemical potential gives access to the compressibility $dn/d\mu$ for each sublattice, which we interpret as a local density-of-state in the absence of interactions (Fig.~\ref{fig:flat_band}E).

The pronounced peak of compressibility on the $p$-sites is indicative of a flat band located on the $p$-sublattice, whereas the other two bands appear as broader peaks above and below half-filling. The finite width of the peaks is explained by non-zero temperature $T/t = \Tzero$, as obtained by fitting the experimental equation of state to the non-interacting expectation for an ideal Lieb lattice.

\placefigure{f2}
\section*{Staggered magnetization}

According to Lieb's theorem \cite{lieb_two_1989}, lifting the state degeneracy at half-filling ($\mu=0$) with any infinitesimal on-site repulsive interaction $U$ gives rise to a ground state with finite total spin, compatible with a ferrimagnetic alignment. In our experimental system however, the total spin is a conserved quantity determined by the populations and coherences of the two atomic hyperfine states encoding spin. For a SU(2)-symmetric, balanced mixture, the total spin is zero on average, which prevents the observation of a ferrimagnetic phase in the thermodynamic limit with a spontaneously broken order parameter, even in the ground state.

Yet, we can probe the ferrimagnetic response of the system by measuring its spatially-resolved magnetization in the presence of a symmetry-breaking Zeeman field $\mathbf{h}$.
Experimentally, we realize an effective global Zeeman field $h$ by preparing a spin-imbalanced atomic mixture, which is achieved by evaporatively cooling the atomic cloud in a magnetic field gradient and at low bias fields. This results in up to a 3:1 ratio in the two spin populations prior to loading into the optical lattice, and a tunable difference $\mu_\uparrow-\mu_\downarrow = 2 h$ between the chemical potentials of both spins after loading (\cite{supplementary} and \cite{brown_spin-imbalance_2017}).
At half-filling and at finite interaction $U/t = \usix$ comparable to the bandwidth $W/t = 4\sqrt{2}$ of the Lieb lattice, we investigate how a global Zeeman field affects magnetic order through measurements of the local magnetization $\langle \hat{S}_z \rangle = n_\uparrow - n_\downarrow$, defined as the projection of the spin operator $\vec{S}$ along the axis $z$ parallel to the Zeeman field, and expressed from average densities $n_\sigma$ of each spin $\sigma$ on a given site.

As shown in Fig.~\ref{fig:vs_magnetization}A-C, we observe the signatures of a ferrimagnetic state when increasing the Zeeman field $h$ as positive magnetic moments $\langle S_z \rangle$ on the $p$-sites of the Lieb lattice (circles), and negative moments on $d$-sites (squares). Their average magnitudes are furthermore different (Fig.~\ref{fig:vs_magnetization}D), and their antialignment persists up to a field $h/t \sim 0.3$, after which the magnetization of $d$-sublattice turns positive (Fig.~\ref{fig:vs_magnetization}D, blue points). The $h$-dependence of the magnetization is consistent with Determinant Quantum Monte Carlo (DQMC) simulations performed in the range of temperatures achieved experimentally $T/t = 0.3 - 0.4$ (\cite{supplementary} and Fig.~S1. See also Fig.~S9 and Sec.~E5 for temperature dependence of the $p$- and $d$-site asymmetry).

We interpret the non-monotonic behavior of the $d$-sublattice magnetization as a competition between polarization induced by the symmetry-breaking field $\mathbf{h}$, and antiferromagnetic correlations among fermions in the repulsive Hubbard model at half-filling \cite{auerbach_interacting_2012}. These spin fluctuations around the average local magnetization are quantified by the connected spin correlator $\langle S_z^i S_z^j \rangle_\text{c} = \langle \hat{S}_z^i \hat{S}_z^j \rangle - \langle \hat{S}_z^i \rangle \langle \hat{S}_z^j \rangle$ between two sites $i, j$ (Fig.~\ref{fig:vs_magnetization}E). The measured correlators are negative for bonds linking $p$- and $d$-sites (bond length $|\mathbf{r}| = |\mathbf{r}_i-\mathbf{r}_j| = 1$ and $3$), and positive between sites belonging to the same sublattice ($|\mathbf{r}| = \sqrt{2}$ and $2$). 
This confirms that the spin fluctuations are quantum and have an antiferromagnetic character upon applying a Zeeman field, in addition to the magnetization being anti-aligned with a finite net moment.

The staggered magnetization $\langle S_z \rangle$ of the Lieb lattice starkly contrasts with the uniform magnetization visible on the surrounding square lattice (Fig.~\ref{fig:vs_magnetization}D, grey points), in which antiferromagnetic correlations are expected to be favored in the orthogonal plane \cite{brown_spin-imbalance_2017}. Despite their different magnetizations $\langle S_z \rangle$, we find that the spin fluctuations $\langle S_z S_z \rangle_c$ in the Lieb and square lattices experimentally exhibit similar correlation lengths, $\xi \sim 0.8$ at $h/t = \hd$ (Fig.~S5). This correlation length is furthermore associated with a finite decay of the Lieb staggered magnetization $\langle S_z \rangle$ into the square lattice with a comparable penetration length (Fig.~S6).

\placefigure{f3}

\section*{Crossover with interactions}

A defining feature of Lieb's ferrimagnetism is its persistence at any interaction strength $U > 0$. Experimentally, we explore the crossover from a metallic to an insulating regime by tuning the onsite interaction energy $U$ using a Fano-Feshbach resonance. 
To confirm the formation of a Mott insulator at half-filling, we first measure the local compressibility $dn/d\mu$ at zero net magnetization through a procedure analogous to Fig.~\ref{fig:flat_band}D and E.
For an interaction $U/t = 6.6(1) \approx W/t$ comparable to the bandwidth (Fig.~\ref{fig:vs_interaction}B), a dip in the compressibility at half-filling signals the onset of the insulating regime. 
The compressibility further gets split at $U/t = \uten$ by an energy gap comparable to $U$ (Fig.~\ref{fig:vs_interaction}C), indicating a Mott insulating state. 
Intriguingly, the compressibility below half-filling (chemical potentials $-10 \leq \mu/t \leq -2$ in Fig.~\ref{fig:vs_interaction}C) resembles the noninteracting compressibility over the entire density range ($-4 \leq \mu/t \leq 4$ in Fig.~\ref{fig:vs_interaction}E). 
We find good agreement between the measured local compressibility and DQMC simulations, as well as Finite Temperature Lanczos Method (FTLM) simulations performed in the range of temperatures achieved experimentally $T/t=0.3-0.5$.
This doubling of the compressibility curve may signal a mapping of the $1/4$-filled, spin-$1/2$ system in the hardcore fermion limit $U \rightarrow \infty$ to a $1/2$-filled, spinless fermionic system. 
A possible explanation is the emergence of a fully polarized ferromagnetic phase \cite{supplementary,gouveia_dias_2015_meanfield1, gouveia_dias_2016_meanfield2}. Our theoretical studies \cite{Nikolaenko_Bonetti_25} have instead led to a remarkable possibility: fractionalization of the spinful fermions into emergent spinless fermions and spinful bosons, with the emergent particles carrying gauge charges under an emergent $\mathbb{Z}_2$ gauge field. This fractionalized structure is the same as that in gapped, insulating spin models like the toric code, but appears here in a gapless metallic system. If established experimentally at lower temperatures, such a state would provide an example of doping-induced fractionalization in a system as simple as the Lieb lattice.

At half-filling, in the presence of a nonzero Zeeman field $h/t \sim 0.2$ applied via a fixed square lattice magnetization $\langle S_z \rangle \sim 0.2$, we observe a robust difference of the magnetization between $p$- and $d$-sites across interactions $U/t = \uzero$ to $\uten$ (Fig.~\ref{fig:vs_interaction}D), akin to the ferrimagnetic pattern in Fig.~\ref{fig:vs_magnetization}B and C. 
In the weakly interacting regime $U \ll W$, a large density-of-state at half-filling is associated with a large magnetic susceptibility; therefore, fixing the spin magnetization to a non-zero value leads to a complete spin polarization of the flat band, \emph{even} in the absence of interactions $U/t = 0$ (Fig.~\ref{fig:vs_interaction}F). 
The full polarization of the flat band and partial polarization of the dispersive bands translate to positive magnetic moments on the $p$-sublattice and a much weaker polarization of the $d$-sublattice.

In the opposite, strongly interacting regime $U \gg W$, the Hubbard model at half-filling maps to an isotropic Heisenberg model with antiferromagnetic couplings set by the superexchange energy $J = 4 t^2 / U$ (Fig.~\ref{fig:vs_interaction}G). 
A bipartite lattice then favors N\'eel ordering, which results in a finite magnetization when the sublattices contain different numbers of sites as in the Lieb geometry \cite{lieb_two_1989,noda_ferromagnetism_2009}. 
As a result the interacting Lieb lattice has opposite signs for the magnetic susceptibility on the $p$- and $d$-sites at temperatures below $J$ (Figs.~S9 and S10). 
Experimentally, we observe a decrease in the magnitude of the staggered magnetization with $U$, which results from the decrease in the superexchange energy $J$ relative to the temperature $T$ (Fig.~S1). 
Imperfections in the decorated potential furthermore leads to an energy offset between the $p$- and $d$-sublattices in this particular dataset (\cite{supplementary}, Fig.~S4 and Fig.~S7), to which ferrimagnetism is however expected to be robust \cite{noda_ferromagnetism_2009}.

Across the experimental interaction range, we measure a connected spin correlation function $\langle S_z S_z \rangle_\text{c}$ consistent with antiferromagnetic spin fluctuations between nearest neighbors (Fig.~\ref{fig:vs_interaction}E, closed symbols). 
By performing a $\pi/2$ radio-frequency pulse to rotate the measurement basis prior to imaging \cite{supplementary}, we further measure the correlation function $\langle S_x S_x \rangle_\text{c}$ along the $x$-direction orthogonal to the Zeeman field (Fig.~\ref{fig:vs_interaction}E, open symbols). 
The measured anisotropy between $z$- and $x$-basis measurements indicates a small enhancement of the antiferromagnetic spin fluctuations in the plane orthogonal to the Zeeman field, similar to previous observations in the square Hubbard model \cite{brown_spin-imbalance_2017}.

\placefigure{f4}

\section*{Square to Lieb geometries}

Using a repulsive potential pattern to decorate individual lattice sites provides a way to continuously tune the unit cell from a Lieb to a square geometry. By decreasing the potential offset $\Delta$ on the decorated sites, we can increase their local density and restore direct tunneling to its neighbors while ensuring that the $p$- and $d$-sites are half-filled.

We observe a smooth evolution of the local magnetization $\langle S_z^p \rangle$ and $\langle S_z^d \rangle$ towards the total magnetization $\langle S_z^\text{square} \rangle = 0.2$ as the decorated potential $\Delta$ is tuned down from $\DeltaFigOffset t$ to zero (Fig.~\ref{fig:vs_offset}A). 
As a result, the $d$-site magnetic susceptibility goes from negative to positive at $\Delta \approx 5t$ for the experimentally realized temperature of $T/t \sim 0.4$. 
We observe a non-monotonic dependence on $\Delta$ of the spin correlations $\langle S_i S_i\rangle_\text{c}$ between nearest-neighbors (Fig.~\ref{fig:vs_offset}B). 
This behavior is reproduced by DMQC simulations and can be explained by correlated tunneling of doublons on the $p$-sites to their two neighboring decorated sites, a process that is resonant at potentials $\Delta = U/2 \approx 3t$.

\section*{Discussion and Outlook}

Our observation of a ferrimagnetic magnetization pattern in response to a weak symmetry-breaking field at finite temperature hints at the existence of a ferrimagnetic phase at vanishing Zeeman field and temperature.
Furthermore, our work calls for a better understanding of the interplay between finite temperature and finite magnetization in multiband models.
In a square lattice, anisotropy introduced by the presence of a Zeeman field is expected to lead to canted antiferromagnetic order at low-temperatures \cite{brown_spin-imbalance_2017}. In a Lieb lattice, Curie-Weiss calculations suggest a staggered spin order parallel to the magnetization which persists to temperatures higher than the canted AFM phase, as well as a smooth crossover from parallel to canted order away from a total magnetization of $\langle S_z \rangle = 1/3$ (\cite{supplementary}, Fig.~S8). Experimentally, our measurements of spin correlations perpendicular to the magnetization $\langle S_x S_x \rangle_\text{c}$ between nearest neighbors highlight the anisotropy between $x$ and $z$ (Fig.~\ref{fig:vs_offset}B).

Future experimental work could open the way to observing magnetic phases with spontaneous symmetry breaking in a SU(2)-symmetric, balanced mixture, for example in multilayer systems \cite{noda_magnetism_2015} which have been recently realized \cite{koepsell_robust_2020,gall_competing_2021}. Further progress on lowering experimental temperatures could help identify a phase transition between parallel and canted order which may occur at half-filling, as suggested by numerical DQMC studies of the attractive Lieb-Hubbard model at finite doping \cite{iglovikov_superconducting_2014}. Away from half-filling, van Hove singularities at densities $n = 1/3$ and $n = 5/3$ may favor itinerant ferromagnetism \cite{auerbach_interacting_2012,xu_frustration-_2023} and could further give rise to a rich phase diagram, as suggested by Finite Temperature Lanczos Method (FTLM) simulations in Figs.~S9 and S10.

Exploring attractive interactions in a Lieb lattice would enable the study of transport and fermionic superfluidity in flat bands \cite{huber_bose_2010,goldman_topological_2011,peotta_superfluidity_2015,julku_geometric_2016,huhtinen_revisiting_2022}. 
Tunable optical lattices \cite{xu_frustration-_2023,lebrat_observation_2024} open exciting prospects for the realization of flat-band geometries central to quantum spin liquids, such as Kagome lattices that can be readily realized by decorating a triangular lattice \cite{yamada_mott_2011}. 
Furthermore, the coexistence of dispersive bands and a flat band with localized magnetic moments calls for a Kondo-lattice type description, similarly to recent work on twisted bilayer graphene \cite{song_TBG_2022, hu_TBG_2023}. 
Such a system with site-resolved control of the Hubbard Hamiltonian would be an ideal setting to explore Kondo and heavy fermion physics \cite{coleman_kondo_lattice, Wirth2016_HeavyFermionReview}.
These capabilities would ultimately help investigate the mapping between the single-band Hubbard Hamiltonian and multi-band models of strongly correlated materials with tunable charge-transfer energy  
\cite{zhang_effective_1988, kowalski_oxygen_2021,simons_absence_2020,jiang_density_2023},
and elucidate the conditions for the emergence of high-temperature superconductivity.

\section*{Acknowledgments}
We thank Annabelle Bohrdt, Nigel Cooper, Fabian Grusdt, Ehsan Khatami, Richard T. Scalettar, and Aaron W. Young for insightful discussions and comments on the manuscript.

\section*{Funding}
We acknowledge support from the Gordon and Betty Moore Foundation, Grant No.~GBMF-11521;
National Science Foundation (NSF) Grants Nos.~PHY-1734011, OAC-1934598 and OAC-2118310;
ONR Grant No.~N00014-18-1-2863;
the Department of Energy, QSA Lawrence Berkeley Lab award No.~DE-AC02-05CH11231;
QuEra grant No.~A44440; ARO/AFOSR/ONR DURIP Grants Nos.~W911NF-20-1-0104 and W911NF-20-1-0163;
the Swiss National Science Foundation and the Max Planck/Harvard Research Center for Quantum Optics (M.L.); 
the NSF Graduate Research Fellowship Program (A.K. and L.H.K.);
the AWS Generation Q Fund at the Harvard Quantum Initiative (Y.G.);
the NSF Grant No.~DMR-2245246 and the Simons Collaboration on Ultra-Quantum Matter which is a grant from the Simons Foundation (651440, S.S.).

\section*{Author contributions}
M.L., A.K., L.H.K., M.X. and Y.G. performed the experiment and analyzed the data. The numerical simulations were performed by A.K. (FTLM), M.L. (DQMC and Curie-Weiss), M.X. (DQMC) and A.N. and P.M.B. (Hartree-Fock). A.N., P.M.B. and S.S. developed the theoretical interpretation of the doped data. M.G. supervised the study. All authors contributed to the interpretation of the results and production of the paper.

\section*{Competing interests}
M.G. is co-founder and shareholder of QuEra Computing.

\begin{dfigure*}{f1}
    \centering
    \noindent
    \includegraphics[width=\linewidth]{"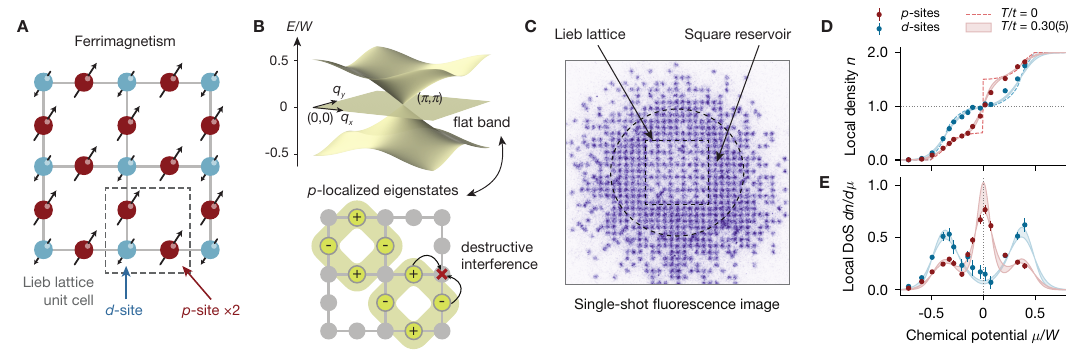"}
    \caption{\textbf{Ferrimagnetism and flat band in a Lieb lattice.}
    (\textbf{A}) Ferrimagnetism refers to a magnetic state with anti-aligned magnetic moments of different magnitudes, resulting in a non-zero total spin. The Lieb lattice features a non-trivial unit cell with two doubly-coordinated $p$-sites, and one fourfold-coordinated $d$-site. Its ground state is a paradigmatic example of ferrimagnetism \cite{lieb_two_1989}, for a half-filled Hubbard model with repulsive on-site interaction $U > 0$.
    (\textbf{B}) This lattice features a large eigenstate degeneracy, lifted by infinitesimal interactions. It is evident as a dispersionless energy band, or equivalently, as degenerate eigenstates spatially localized on the $p$-sites of a square plaquette. 
    (\textbf{C}) We realize the Fermi-Hubbard model on $5\times5$ unit cells of a Lieb lattice with lithium-6 atoms and a repulsive potential decorating every fourth site of a square optical lattice. The non-decorated surrounding lattice acts as a bath and thermometer.
    (\textbf{D}) We probe the non-interacting ($U/t=\uzero$) equation of state by controlling the total atom number after loading which allows us to tune the chemical potential $\mu$, here normalized by the total bandwidth $W = 4\sqrt{2} t$. 
    The symmetry of the density on the $p$- and $d$- sublattices around half-filling highlights the particle-hole symmetry of the band structure. 
    (\textbf{E}) We convert sublattice densities into local density-of-states (DoS) through a numerical derivative with respect to $\mu$. The flat band appears as a pronounced peak at half-filling on the $p$-sublattice broadened by finite temperature, here fitted to $T/t = \Tzero$. Shaded bands indicate non-interacting calculations for $T/t = 0.25-0.35$. Here and in the following, error bars indicate the s.e.m.}
    \label{fig:flat_band}
\end{dfigure*}

\begin{dfigure*}{f2}
    \centering
    \noindent
    \includegraphics[width=\linewidth]{"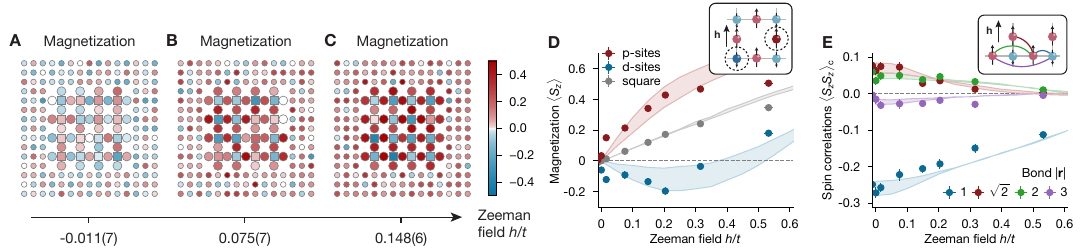"}
    \caption{\textbf{Ferrimagnetic susceptibility with applied Zeeman field.}
    (\textbf{A}-\textbf{C}) 
    Preparing a spin-imbalanced mixture leads to a weak effective Zeeman field $h$ that tends to align the magnetization along the $z$ axis, which we measure through spin-resolved imaging. 
    At interactions $U/t = 6$ comparable to the total bandwidth $W$, increasing $h$ leads to the appearance of antialigned magnetization $\langle S_z \rangle$ on the $p$-sites (circles) and $d$-sites (squares) of the Lieb lattice, whereas the reservoir sites do not show a staggered magnetization (smaller circles).
    (\textbf{D}) 
    The response of the local magnetization to a small Zeeman field reveals the ferrimagnetic susceptibility of the state. 
    The local magnetization $\langle S_z \rangle$ of the $p$- and $d$-sites (red and blue circles) has different amplitudes and opposite signs up to a Zeeman field $h/t = \hd$. 
    At higher Zeeman field $h/t = \hmax$, both sublattices become polarized, $\langle S_z \rangle > 0$, but have different amplitudes. 
    By contrast, magnetization on the square lattice increases monotonically with $h/t$ (grey points).
    Shaded bands indicate Determinant Quantum Monte Carlo (DQMC) simulations performed for a representative range of experimental temperatures $T/t = 0.3$ to $0.4$ (Fig.~S1). 
    (\textbf{E}) 
    Spin fluctuations around the average magnetization are quantified by the connected, two-point spin correlation function $\langle S_z S_z \rangle_\text{c}$, in which local magnetic moments appear as a disconnected part and are subtracted off.
    Spin correlations are negative between $p$- and $d$-sites (bond length $|\mathbf{r}| = 1$ and $3$), and positive between pairs of sites belonging the same sublattice ($|\mathbf{r}| = \sqrt{2}$ and $2$), indicating antiferromagnetic fluctuations. 
    The associated spin correlation length is comparable to the square lattice (Fig.~S5).
    }
    \label{fig:vs_magnetization}
\end{dfigure*}

\begin{dfigure}{f3}
    \centering
    \noindent
    \includegraphics[width=\linewidth]{"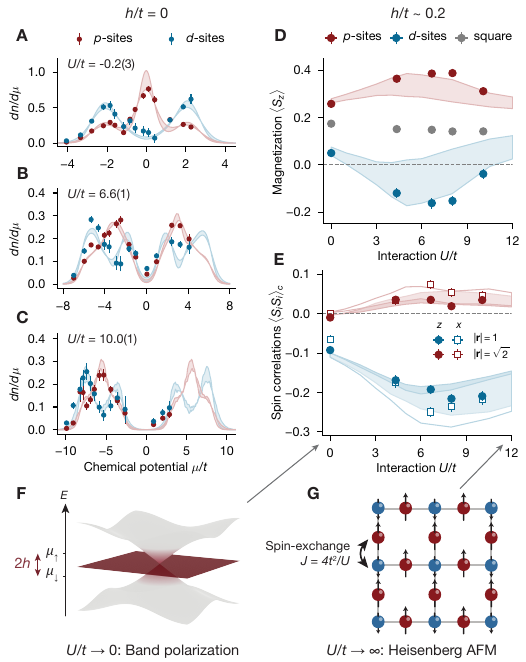"}
    \caption{\textbf{Crossover with interactions.}
     Local compressibility $dn/d\mu$ as a function of interaction strength (\textbf{A}) $U/t = \uzero$ (reproduced from Fig.~\ref{fig:flat_band}E), (\textbf{B}) $U/t = \usix$ and (\textbf{C}) $U/t = \uten$, in the absence of global magnetization. The band structure visible at $U/t = 0$ becomes split by a Mott gap of energy $U$, associated with a vanishing compressibility at half-filling. The similarity between the compressibility without interactions and the compressibility over the hole-doped half range $\mu < 0$ at $U/t = 10$ suggests fractionalization to spinless fermions and spinful bosons which is studied in Ref.~\cite{Nikolaenko_Bonetti_25}.
     (\textbf{D}) At half-filling and non-zero Zeeman field $h/t \sim 0.2$, the magnetization on the $p$-, $d$- and square sites confirm the ferrimagnetic character of the Lieb lattice over an interaction range $U/t = 0$ to $10$ covering the metallic and insulating regimes. (\textbf{E}) The connected spin correlation function $\langle S_i S_i \rangle_\text{c}$ is anisotropic along axes $i = z, x$, indicating a small enhancement of the antiferromagnetic spin fluctuations in the plane orthogonal to the Zeeman field. In panels b--e, shaded bands indicate DQMC simulations at $T/t = 0.3-0.5$.
     (\textbf{F}) Without interactions, a Zeeman field strongly polarizes the flat band, leading to larger positive magnetizations on the $p$-sublattice than the $d$-sublattice. (\textbf{G}) In the Heisenberg limit, $U/t \rightarrow \infty$, spin-exchange $J = 4t^2/U$ leads to antiferromagnetic correlations on the Lieb lattice.
    }
    \label{fig:vs_interaction}
\end{dfigure}

\begin{dfigure}{f4}
    \centering
    \noindent
    \includegraphics[width=\linewidth]{"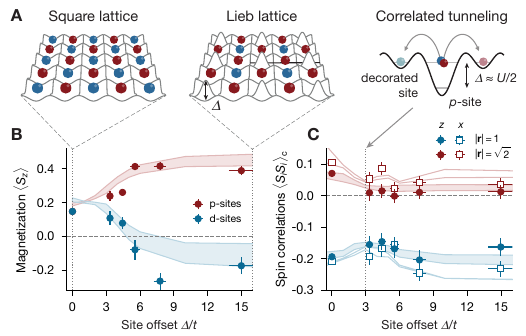"}
    \caption{\textbf{Tunable unit cell and correlated hopping.}
    (\textbf{A})~Tuning the potential offset $\Delta$ on the decorated lattice sites allows to continuously change the lattice from a  square to a Lieb geometry (right). (\textbf{B})~Maintaining the non-decorated sites at half-filling and at interaction strength $U/t = \usix$, we observe a convergence of the magnetization on the $p$- and $d$-sites (excluding decorated sites) towards the fixed average magnetization $\langle S_z \rangle \sim 0.2$ as the offset $\Delta$ is decreased. (\textbf{C})~The connected spin correlation function shows a broad peak around $\Delta = 3t \approx U/2$, hinting at a resonant process involving the correlated tunneling of two spins on a $p$-site to its two adjacent decorated sites. Bands: DQMC simulations performed in a temperature range $T/t = 0.3 - 0.4$ with average magnetization $\langle S_z \rangle = 0.2$.
    }
    \label{fig:vs_offset}
\end{dfigure}

\setcounter{figure}{0}
\renewcommand{\thefigure}{S\arabic{figure}} 

\clearpage

\section*{Supplementary Materials}

\subsection{Experimental sequence} \label{sec:exp_sequence}

We prepare an ultracold sample of $^6$Li in a mixture of two hyperfine states, labeled as $\uparrow$ and $\downarrow$ respectively. For most data, we use the lowest state $\ket{\uparrow} = \ket{F=1/2, m_F=1/2}$ and the third lowest state $\ket{\downarrow} = \ket{F=3/2, m_F=-3/2}$, excepted for the spin-balanced data at $U/t = 6$ in Fig.~2, where we use the lowest and second lowest states $\ket{\downarrow} = \ket{F=1/2, m_F=-1/2}$.
To prepare a spin-imbalanced sample, we perform evaporative cooling in an optical dipole trap with a magnetic field gradient before lattice loading, which leads to a differential trap depth due to the different magnetic moments of the two hyperfine states.
The parameters of this last evaporation step determine the total magnetization of the cloud, defined as $M_z = (N_\uparrow-N_\downarrow)/(N_\uparrow+N_\downarrow)$, where $N_\uparrow$ and $N_\downarrow$ are the populations of the two states.
We obtain spin-balanced mixtures at evaporation magnetic fields $B = \SI{590}{\gauss}$, deep in the Paschen-Back regime, and maximally imbalanced mixtures with $M = 0.36$ at low fields $B = \SI{40}{\gauss}$, where the difference of the magnetic moments is larger than 50\%.

After evaporation, the magnetic field is ramped to its final value between $B = \SI{568}{\gauss}$ for non-interacting data (vanishing $s$-wave scattering length $a_s = 0$) and $B = \SI{598}{\gauss}$ for $U/t = 12$ data ($a_s = 535\,a_0$, where $a_0$ is the Bohr radius). 
We then simultaneously ramp up the lattice depth of the square optical lattice with $a=\SI{805}{\nano\meter}$ \cite{xu_frustration-_2023} and the repulsive potential used to decorate the lattice in $\SI{80}{\milli\second}$.

Similar to \cite{lebrat_observation_2024}, we perform single-site resolved imaging after splitting every lattice site into two sites using a super-lattice which allows measurement of the full density distribution (0, 1, and 2 atoms).
The splitting of the super-lattice relies on repulsive interactions between two atoms on the same site and hence before performing the splitting step, we ramp the magnetic field to have repulsive interactions.
For measuring spin correlations in the usual $z$ basis, we take two additional kinds of snapshots to image (i) only the $\ket{\uparrow}$ atoms or (ii) only the $\ket{\downarrow}$ atoms by using resonant laser pulses to blow out atoms of the other spin state \cite{parsons_site-resolved_2016}.
In particular, in order to image $\ket{\uparrow}$ ($\ket{\downarrow}$) atoms, we first perform an interaction-resolved radio-frequency Landau-Zener (LZ) sweep to transfer only the singles in state $\ket{\downarrow}$ ($\ket{\uparrow}$) to state $\ket{2} = \ket{F=1/2, m_F=-1/2}$, followed by a resonant laser beam to blowout state $\ket{2}$ atoms. 
This interaction resolved LZ step ensures that doublons survive the blowout beam and can be imaged as two atoms after the super-lattice splitting.
For measuring spin correlations in the $x$ basis, before shining the resonant laser beam to blowout state $\ket{2}$, we perform an additional radio-frequency $\pi/2$ Rabi pulse between states $\ket{\uparrow} \leftrightarrow \ket{2}$ ($\ket{\downarrow} \leftrightarrow \ket{2}$).  
These snapshots can be combined to compute average two-point spin correlations \cite{parsons_site-resolved_2016}.

The repulsive potential used to decorate the square lattice is created by blue-detuned, $\SI{650}{\nano\meter}$ beam, whose intensity is spatially modulated by a Digital Micromirror Device (DMD) which is optically conjugated with the atomic plane. To maintain the relative alignment between the DMD projected repulsive potential and the optical lattice, we perform active feedback on the DMD pattern \cite{ji_thesis_2020}. After every 20 experimental snapshots ($\sim$ 8 minutes), we run one calibration shot in which we use the DMD to project a sinusoidal calibration pattern on the atoms that allows us to fit the relative position of the DMD pixels and the optical lattice sites. We then update the pattern projected on the DMD with the new fitted position. We estimate the drift of the relative position between DMD and lattice sites to be less than about one tenth of a lattice site over 200 experimental shots.

\subsection{Calibrations} \label{sec:calibrations}
\subsubsection{Temperature $T$}
We use the square lattice reservoir for thermometry of the different datasets. By comparing the experimentally measured nearest-neighbor and next-nearest neighbor spin correlations at half-filling from the square reservoir to DQMC simulations, we can compute the temperature of the reservoir and hence the temperature of the Lieb lattice (Fig.~\ref{fig:temperature}, \ref{fig:temperature_vs_filling}).

\placefigure[!t]{s1}

\placefigure[!t]{s1b}

\placefigure[!t]{offset_calibration}

\subsubsection{Chemical potential $\mu$} \label{sec:chem_pot}
In our experiment, we can control the total atom number as an experimental knob but not the absolute chemical potential. 
However, if we know the equation of state for the square lattice, we can infer the absolute chemical potential in the square region from the experimentally measured density by inverting the equation of state. 
We can then relate the chemical potential of the Lieb lattice to that of the square region because the two systems are in equilibrium.  
For the case of no interactions, we can compute the equation of state exactly for the square lattice as well as the Lieb lattice. 
For Fig.~1, we use the experimentally measured densities in the square reservoir to infer the chemical potential $\mu_{\text{square}}(N_\text{tot})$ as we change the total atom number $N_\text{tot}$. 
We also find the chemical potential for the Lieb region $\mu_{\text{Lieb}}(N_{\text{tot}})$ assuming the equation of state for an ideal Lieb lattice (no disorder, periodic boundaries, no finite size effects). 
We find that the two chemical potentials are simply related by an offset that is atom number independent, 
\begin{equation} \label{eq:chemical_potential_lieb_square}
    \mu_{\text{Lieb}}(N_\text{tot}) = \mu_{\text{square}}(N_\text{tot}) + \delta.
\end{equation} 
The fitted offset $\delta = 0.81(2)$ is explained by residual blue-detuned light from the DMD potential being spread to the populated sites (see Section~\ref{sec:disorder}).
For the interacting case in Fig.~3A-C, we use the measured density in the square reservoir and a numerically obtained equation of state for the square lattice via Numerical Linked Cluster Expansion (NLCE) simulations \cite{khatami_thermodynamics_2011} to compute $\mu_{\text{square}}$ and hence $\mu_{\text{Lieb}}$ via Eq.~(\ref{eq:chemical_potential_lieb_square}).

\subsubsection{Hubbard parameters $t, U$} \label{sec:hubbard_cal}
The tunneling amplitudes of the underlying square lattice are obtained as in \cite{xu_frustration-_2023}, and are equal to $t_x = \SI{0.36(1)}{\kilo\hertz}$ and $t_y = \SI{0.33(1)}{\kilo\hertz}$ between nearest-neighbor sites.
The onsite interaction energy $U$ is calibrated by measuring the experimental single-occupancy probability in the square reservoir, averaged over sites with density within $\pm0.03$ from half-filling, and comparing it to NLCE data \cite{khatami_thermodynamics_2011} at $T/t = 0.4$ and DQMC simulations.

\subsubsection{Potential offset $\Delta$}

The potential offset on the decorated site $\Delta$ is obtained from the dataset of Fig.~4, using a least-squares fit of the decorated-site densities against the power of the DMD beam. The fit function is obtained from numerical DQMC simulations performed as a function of $\Delta/t$ at $U/t = 6$ and $T/t = 0.3$ (Fig.~\ref{fig:calibrate_offset}), with a linear scaling factor relating DMD power and $\Delta/t$ that acts as a single free parameter. At the maximal DMD power used to realize the Lieb lattice, the fitted potential offset averaged over the full $5\times 5$ array of repulsive tweezers is equal to $\Delta/t = \DeltaMain$, where the error indicates the spatial variation of $\Delta$, expressed as a standard deviation, stemming from the envelope of the beam illuminating the DMD.

The three lowest bands of the Lieb-lattice band structure, which are relevant for Figs.~1 to 3, are weakly sensitive to this relative change in $\Delta$ (Section~\ref{sec:band_structure}). It however leads to absolute changes in the gap $\sim \Delta$ to the fourth band, and inhomogeneities of the density on the decorated sites at intermediate offset potentials in Fig.~4 over the full $5\times5$ array. To reduce this decorated density inhomogeneities, we restrict the analysis of Fig.~4 to a smaller region of interest including $3\times3$ repulsive tweezers with maximal average $\Delta/t = \DeltaFigOffset$ (Fig.~\ref{fig:calibrate_offset}).

\placefigure[!t]{disorder}

\subsection{Uniformity of the Lieb-lattice sites} \label{sec:disorder}

The finite resolution of the DMD imaging system leads the blue-detuned spots projected on the decorated sites to bleed over the neighboring, populated sites. This bleedover can be quantified by the ratio of the repulsive potential on the $p$- or $d$-sites $\Delta_{p,d}$ to the potential on the decorated sites $\Delta$, which we estimate by measuring the optical powers required to deplete the decorated sites and the Lieb sites in a deep optical lattice. Although we expect the imaging point-spread-function to be narrower than the lattice spacing $a=\SI{805}{\nano\meter}$, resulting in a bleedover ratio better than 1:50 when decorating well-separated sites, interference between repulsive spots on a tightly-spaced array with period $2a$ enhances this ratio to about 1:10. For a perfect alignment of the decorated potential onto the square lattice, the residual potential is furthermore made equal on the $p$- and $d$-sites, $\Delta_{d} = \Delta_{p}$, by fine tuning the width of the repulsive spots through the DMD pattern. The estimated bleedover ratio of 1:10 accounts for the fitted energy offset of about $1t$ between the square and Lieb lattice at maximum $\Delta = 11.1(6)t$ reported in Section~\ref{sec:chem_pot}.

Experimental datasets may still exhibit non-uniform on-site potentials due to the spatial profile on the beam illuminating the DMD or imperfect alignment of the DMD onto the optical lattice. We estimate this non-uniformity by plotting the distribution of the on-site densities on all $p_x$-, $p_y$ and $d$-sites (shown in Fig.~\ref{fig:disorder}a, b, and c as left-pointing triangles, right-pointing triangles, and circles, respectively), statistically averaged over the datasets of Fig.~2 to Fig.~3.

The spatially averaged densities on the $p_x$, $p_y$ and $d$ sublattices are consistent with each other and within 5\% for most datasets, with the exception of Fig.~3, which shows a density of about 0.8 on the $d$-sublattice at $U/t = 0$. This density difference can be related to a chemical potential energy difference through the non-interacting equation of state $n(\mu, T)$ on the $p$- and $d$-sites of the Lieb lattice shown in Fig.~\ref{fig:disorder}d. The observed depletion of the $d$-sublattice translates to an energy offset relative to the $p$-sublattice of order $t$, most likely owing to a slight misalignment of the DMD onto the optical lattice in this particular dataset.

The antialigned magnetizations visible in Fig.~3 are however robust to the energy offset between the $p$- and $d$-sites, as described in \cite{noda_ferromagnetism_2009} in the limit where this offset is smaller than the interaction energy $U$. We furthermore note that average densities on the $p_x$- and $p_y$-sites are balanced; equal $p_x$ and $p_y$ potentials preserve the flatness of the second band in a Lieb lattice with a three-site unit cell (Section~\ref{sec:band_structure} and Fig.~\ref{fig:band_structure}b).

\subsection{Correlation length and Lieb-square boundary}

\placefigure[!t]{s4}

\placefigure[!t]{boundary}

We fit an exponential decay to the connected spin correlations $\langle S_z S_z \rangle_c$ in the Lieb lattice and the surrounding square reservoir for the datasets in Figs.~ 2 and 3, see Fig.~\ref{fig:corr_length}. We find a similar correlation length in the Lieb and square regions.

Breaking translation symmetry with the decorated potential also leads to the pinning of local magnetic moments within the Lieb lattice, which are expected to decay in the square lattice. We plot in Fig.~\ref{fig:boundary}a the magnetization $\langle S_z^i \rangle$ across the Lieb-to-square boundary (defined as the dashed square shown in Fig.~1D), averaged over every half-line and half-column of Fig.~2C which spans both Lieb and square lattices and does not include a decorated site. To visualize the penetration depth of the ferrimagnetic moments into the square lattice, we further plot in Fig.~\ref{fig:boundary}b the staggered magnetization at site index $i$
\begin{equation} \label{eq:stagmag}
    \tilde{m}_z^i = (-1)^i (\langle S_z^i \rangle - \langle\bar{S}_z\rangle)
\end{equation}
where $\langle\bar{S}_z\rangle$ is the spatial average of $\langle S_z^i \rangle$ over all sites $i$. The staggered magnetization decays away from the boundary over a couple of sites for the finite-magnetization data of Fig.~2B, C, suggesting that the penetration depth of the staggered magnetization is comparable to the correlation length of the connected spin correlations $\langle S_z S_z \rangle_c$ measured independently.

\placefigure[!t]{bands}

\subsection{Numerical simulations}

\subsubsection{Band structure} \label{sec:band_structure}
The band structure is computed by diagonalizing the tight-binding Hamiltonian on a square lattice with a unit cell including four inequivalent orbitals $d$, $p_x$, $p_y$, $o$ for the sake of generality. This Hamiltonian can be expressed in the basis of Bloch states at quasi-momentum $\textbf{q} = (q_x, q_y)$ (setting in this section the lattice spacing to $a = 1$):
\begin{widetext}
\begin{equation}
\hat{H}_\mathbf{q} = t \begin{pmatrix}
\Delta_d/t & -2 \cos(\pi q_x) & -2 \cos(\pi q_y) & 0 \\
-2 \cos(\pi q_x) & \Delta_{p_x}/t & 0 & -2 \cos(\pi q_y) \\
-2 \cos(\pi q_y) & 0 & \Delta_{p_y}/t & -2 \cos(\pi q_x) \\
0 & -2 \cos(\pi q_y) & -2 \cos(\pi q_x) & \Delta/t
\end{pmatrix}    
\end{equation}
\end{widetext}
where $t$ is tunneling energy and $\Delta_i$ is the potential energy on orbital $i$. In the Lieb limit of a large energy offset on the decorated orbital, $\Delta/t \rightarrow +\infty$, it reduces to a $3 \times 3$ Hamiltonian coupling orbitals $p_x \leftrightarrow d$ and $d \leftrightarrow p_y$. 

We illustrate in Fig.~\ref{fig:band_structure} the effect of on-site offsets on the band structure. A large but finite offset $\Delta$ on the decorated site breaks particle-hole symmetry, resulting in a fourth band at energy $E = \Delta$ and a slight broadening of the flat band, amounting to a bandwidth of about $0.3t$ for an offset of $\Delta = 12 t$ (Fig.~\ref{fig:band_structure}a). In the presence of a non-zero, balanced offset $\Delta_{p_x} = \Delta_{p_x} \equiv \Delta_p$, the Hamiltonian still admits an eigenstate with no overlap on the $d$ orbital and with energy $\Delta_p$, equivalent to the dark state of an atomic three-level lambda system. This eigenstate is independent of $\mathbf{q}$ and leads to a flat band. Particle-hole symmetry is also broken, with a gap opening between the second and the third band at the Dirac point $\mathbf{q}_M = (\pi, \pi)$ equal to $|\Delta_d-\Delta_p|$ (Fig.~\ref{fig:band_structure}b, $\Delta_d = t$).

\subsubsection{Determinant Quantum Monte Carlo (DQMC)} \label{subsec:dqmc}
We use the QUEST package \cite{varney_QUEST_2009} to perform unbiased simulations of the Fermi-Hubbard model on a 48-site Lieb lattice ($4\times 4$ unit cells) using the Determinant Quantum Monte Carlo (DQMC) algorithm. For each run, we ensure convergence by using from $2000$ to $5000$ warmup passes, from $12000$ to $30000$ measurement passes and a Trotter step size $t d\tau$ from $0.04$ to $0.01$.
In the Lieb lattice at half-filling and any Zeeman field, there is no sign problem for the Monte Carlo sampling and hence we can reach low temperatures with converged results. Away from half-filling, the sign problem prevents us from reaching temperatures below $T/t \sim 0.2$ on a desktop computer for the 48-site system size.

\subsubsection{Finite Temperature Lanczos Method (FTLM) simulations} \label{subsec:ftlm}
We use the Finite Temperature Lanczos Method (FTLM) \cite{lanczos1950,prelovsek2017} to compute thermal expectation values of observables on a 12-site Lieb lattice ($2\times2$ unit cells).
As previously described in Ref.~\cite{lebrat_observation_2024}, we use an order M = 75 Lanczos decomposition which is typically enough to converge the ground state energy, and use 500 random seeds in each of the symmetry sectors.

When performing exact diagonalization, we make use of certain symmetries of the Hamiltonian (namely particle number $N$ conservation and $S^z_{\text{tot}}$ conservation) to reduce the dimensionality of the matrix representation of operators.
Within each symmetry sector, we obtain the partition function $Z_\beta(N, S^z_{\text{tot}})$ and observables, say $A_\beta(N, S^z_{\text{tot}})$ using FTLM for inverse temperature $\beta = 1/T$. 
Thermal expectation values are then obtained by combining the results from different symmetry sectors by weighting them with the appropriate statistical weights.
In our previous work, we worked in the canonical ensemble (fixed particle number) and hence took thermal averages over all possible magnetization sectors $S^z_{tot} = N_\uparrow-N_\downarrow$. 
In this work, we utilize the grand canonical ensemble (fixed chemical potential) by averaging over different particle number sectors, and magnetization sectors with the correct grand potential weights. 
In addition, we can work with a finite Zeeman field by adding a differential chemical potential $\Delta \mu$ that couples to $S^z_{\text{tot}}$ in the grand potential. 
The grand partition function $Z_\beta(\mu, \Delta \mu)$ is then given by:
\begin{equation}
    Z_\beta(\mu, \Delta \mu) = \sum_N{Z_\beta(N, S^z_{\text{tot}}) e^{\beta (\mu + \frac{U}{2}) N} e^{\beta \Delta \mu S^z_{\text{total}}}}.   
\end{equation}
Observables $A_\beta(\mu,\Delta \mu)$ can be computed as:
\begin{multline}
    A_\beta(\mu,\Delta \mu) = \\
    \frac{1}{Z_\beta(\mu, \Delta \mu)} \sum_N{e^{\beta (\mu+\frac{U}{2}) N} e^{\beta \Delta \mu S^z_{\text{tot}} } Z_\beta(N, S^z_{\text{tot}}) A_\beta(N, S^z_{\text{tot}})}.  
\end{multline}

\subsubsection{Curie-Weiss calculation in Heisenberg model} \label{sec:curie-weiss}

Insights on the phase diagram of a spin-imbalanced Hubbard system in a Lieb lattice can be gained in the large $U/t$ limit by considering a Heisenberg model with Hamiltonian
\begin{equation}
\hat{H} = (J/2) \Sigma_{(i,j)} \hat{\mathbf{S}}_i \cdot \hat{\mathbf{S}}_j - \Sigma_i \mathbf{h} \cdot (\hat{\mathbf{S}}_i - \mathbf{m})
\end{equation}
where $\hat{\mathbf{S}}_i$ is the spin operator on site $i$ and the first sum runs over pairs of nearest-neighbor sites.

Following \cite{koetsier_imbalanced_2010}, the constraint of a finite average magnetization $\langle S_z \rangle = m$ is enforced by adding a Zeeman field $\mathbf{h}$ acting as a Lagrange multiplier.
Under a mean-field approximation, we consider local spin fluctuations around the expectation values of the magnetization on the $p$- and $d$-orbitals, which can be parameterized as $\mathbf{m}_p = \mathbf{m} + (1/3) \mathbf{n}$ and $\mathbf{m}_d = \mathbf{m} - (2/3) \mathbf{n}$, where we define the staggered magnetization $\mathbf{n} = \langle \hat{\mathbf{S}}_p \rangle - \langle \hat{\mathbf{S}}_d \rangle$ under the requirement that there are twice as many $p$-orbitals as $d$-orbitals (Fig.~\ref{fig:curie-weiss}a). The free energy per particle can then be derived as:
\begin{align*}
    f &= \mathbf{h} \cdot \mathbf{m} - \frac{4J}{3} \left(\mathbf{m}+\frac{1}{3}\mathbf{n}\right) \cdot \left(\mathbf{m}-\frac{2}{3}\mathbf{n}\right) \\
    &-\frac{2}{3}{k_B T}{J} \ln \left( 2\cosh \frac{|\mathbf{h}_d|}{2k_B T}\right) \\
    &-\frac{1}{3}{k_B T}{J} \ln \left(2\cosh \frac{|\mathbf{h}_p|}{2k_B T}\right)
\end{align*}
with effective Zeeman fields on the $d$ and $p$ sublattices:
\begin{align*}
    \mathbf{h}_d &= \mathbf{h} - 2J\left(\mathbf{m}-\frac{2}{3}\mathbf{n}\right) \\
    \mathbf{h}_p &= \mathbf{h} - 2J\left(\mathbf{m}+\frac{1}{3}\mathbf{n}\right).
\end{align*}
For a given average magnetization $m$ and temperature $T$, thermal equilibrium is reached at a saddle point of the free energy, for which $f/J$ is maximal with respect to the Zeeman field $\mathbf{h}$ and minimal with respect to the staggered magnetization $\mathbf{n}$.

The phase diagram obtained from numerical optimization and shown in Fig.~\ref{fig:curie-weiss}b indicates the coexistence of two ordered phases: a ferrimagnetic phase in which staggered magnetization $\mathbf{n}$ and global magnetization $\textbf{m}$ are parallel ($\theta = 0$, red), and a canted antiferromagnetic phase in which they are orthogonal (cAFM, $\theta = \pi/2$, blue). Ferrimagnetic order is preferred down to zero temperature at global magnetizations $|\textbf{m}| = 1/3$, at which the $2:1$ number ratio between $p$- and $d$-sites perfectly matches the ratio between up and down spins. Deviations from this magnetization at temperatures $T \lesssim J/2$ lead to a progressive canting of the antiferromagnetic order $\theta > 0$. By contrast, mean-field analysis on a square lattice \cite{koetsier_imbalanced_2010} shows a direct transition from a paramagnetic phase ($|\mathbf{n}| = 0$) to a fully canted AFM phase ($|\mathbf{n}| > 0$, $\theta = \pi/2$), see Fig~\ref{fig:curie-weiss}c.

Quantum fluctuations beyond mean-field are generally expected to weaken spin order and reduce transition temperatures. In particular, antiferromagnetic ordering strictly occurs at $T = 0$ in two dimensions and in the absence of symmetry-breaking field, $\textbf{m} = 0$. The mean-field diagram in Fig.~\ref{fig:curie-weiss}a however suggests that ferrimagnetism should qualitatively survive in the Lieb lattice at temperatures much higher than the canted AFM phase. In the square lattice, the critical temperature for a Kosterlitz-Thouless transition to the canted AFM phase is known to be at most $T_c/t = 0.15$ at $U/t = 4$ in the square Hubbard model \cite{paiva_critical_2004}.

\placefigure[!th]{curie-weiss}

\subsubsection{Additional numerical results}\label{subsec:additional_results}
Here we show numerical results for ferrimagnetism in the Lieb lattice for different interaction strengths.

\placefigure[!th]{ferrimagnetism_numerics}

In Fig.~\ref{fig:ferrimagnetism_numerics}a, b, c, we show the local magnetization as a function of Zeeman field for interaction strengths $U/t=0, 6,$ and $12$.
For zero interactions and very low temperatures, even a small Zeeman field polarizes the p-sites due to the contribution of the flat band. 
At interaction strength $U/t=6$ and $U/t=10.5$, we find good agreement between DQMC and FTLM simulations at temperature of $T/t=0.3$. The FTLM underestimates the $d$-site magnetization for a range of Zeeman fields possibly due to finite size effects. 

In Fig.~\ref{fig:ferrimagnetism_numerics}d, e, f, we show the magnetic susceptibility $\chi$, defined as 
\begin{equation}
    \chi = \frac{d\langle S_{z,i}\rangle}{dh}\Big|_{h=0}, 
\end{equation}
for the $d$- and $p$-sites and their average, as a function of temperature. 
At half-filling, we find a diverging total susceptibility for decreasing temperature indicative of a ferromagnetic state. 
In addition, for the non-interacting case the susceptibility for the $d$-sites is zero at low temperatures, while for the interacting cases we find opposite signs for the susceptibility on the $p$- and $d$-sites (ferrimagnetism).

\placefigure[!th]{s7}

In Fig.~\ref{fig:susceptibility_vs_T_n}, we show the magnetic susceptibility $\chi$ below half-filling for the $d$ and $p$-sites and their average, as a function of temperature $T$ and average density $n$. 
At low temperatures and close to half-filling $n \lesssim 1$, we find a region of ferrimagnetism indicated by the opposite and diverging susceptibilities for the $d$- and $p$-sites. 
For $U/t=6~ (12, 20)$ and $n\sim 0.75~ (0.66, 0.66)$, we find a region with very small susceptibility at low temperatures. 
The statistical error bars in the FTLM simulation become quite significant in this region, and we find disagreement between DQMC and FTLM simulations for these parameters.
We leave the exploration of this quarter-filled state to future work.
At lower densities, we find a ferromagnetic region close to $n\sim 0.33$ for $U/t=6, 12$, and $20$ which may be explained by a Stoner-type mechanism.

\subsubsection{Hartree-Fock calculation in Hubbard model} \label{subsec:hartree-fock}

\placefigure[!th]{hartree_fock}

We reproduce the results of Ref.~\cite{gouveia_dias_2016_meanfield2} by performing a Hartree-Fock calculation on the Lieb-lattice Hubbard model, working with chemical potential instead of fixed particle number. 
In Fig.~\ref{fig:hartree-fock}a, we compare the mean-field energies of the antiferromagnetic, ferromagnetic and paramagnetic state for $U/t=12$. 
Within mean-field, we find that the ferromagnetic state is energetically favored for densities below around $0.9$ (and above around $1.1$). 
Using the magnetization that minimizes the mean-field energy, we obtain an equation of state (Fig.~\ref{fig:hartree-fock}b) which also displays a doubling of the band structure similar to Fig.~3. 

We notice that the mean-field prediction of ferromagnetism is consistent with some ferromagnetic regions in the FTLM calculations of the magnetic susceptibility below half-filling (Section \ref{subsec:additional_results}) but is at odds with the sign structure of the two-point spin correlations versus bond distance obtained with our finite-temperature DQMC and FTLM simulations.
We also note that mean-field calculations tend to incorrectly predict ferromagnetic order in certain Hubbard models, such as the square lattice at finite doping \cite{hirsch_two-dimensional_1985}.
A more detailed study of the doped Lieb-lattice Hubbard model by Hartree-Fock and other methods is presented elsewhere \cite{Nikolaenko_Bonetti_25}.

\begin{dfigure}{s1}
    \centering
    \noindent
    \includegraphics[width=\columnwidth]{"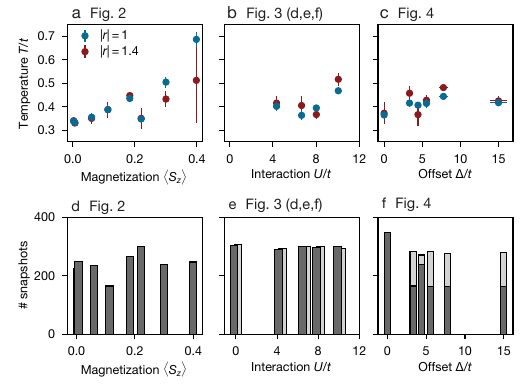"}
    \caption{\textbf{Calibrated temperatures.}
    Temperature calibrated from nearest-neighbor (next nearest-neighbor) spin correlations in the square lattice shown in circles (squares) for datasets: (a) at fixed interaction $U/t = 6$ and varying magnetization $m$ (Fig.~2); (b) at fixed magnetization $\langle S_z \rangle \approx 0.2$ and varying interaction $U/t$ (Fig.~3); (c) at fixed magnetization $\langle S_z \rangle \approx 0.2$, interaction $U/t = 6$ and varying potential on decorated site $\Delta/t$ (Fig.~4).
    (d, e, f) Number of experimental snapshots taken.
    Dark grey bars indicate number of snapshots taken using Z basis measurements and light grey bars indicate X basis measurements.
    \label{fig:temperature}}
\end{dfigure}

\begin{dfigure}{s1b}
    \centering
    \noindent
    \includegraphics[width=\columnwidth]{"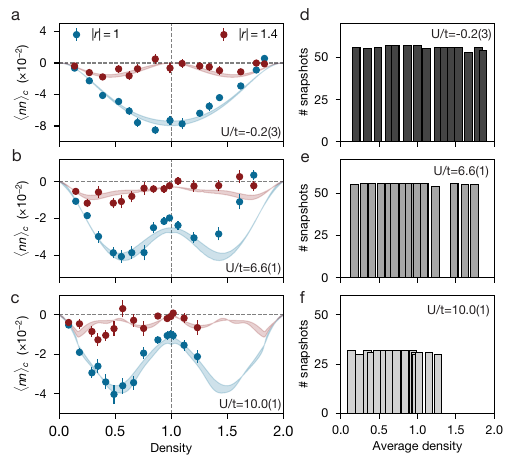"}
    \caption{\textbf{Temperature calibration using density correlations $\langle S_z\rangle=0$.}
    (a,b,c) Connected density-density correlator $\langle n_i n_j \rangle_c$ computed in the square reservoir. Shaded bands indicate numerical results for temperatures in the range (a) $T/t \in [0.2, 0.4]$ for $U/t=0$, (b) $T/t \in [0.3, 0.5]$ for $U/t=6.5$, and $U/t=10.5$.   
    (d,e,f) Number of experimental snapshots taken plotted versus density of a fixed bin in the square reservoir.
    \label{fig:temperature_vs_filling}}
\end{dfigure}

\begin{dfigure}{curie-weiss}
    \centering
    \noindent
    \includegraphics[width=\columnwidth]{"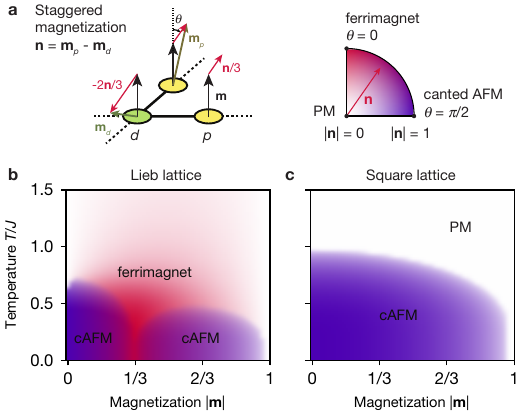"}
    \caption{\textbf{Curie-Weiss mean-field phase diagram.}
    (a) N\'eel order on a bipartite lattice is characterized by a non-zero staggered magnetization $\textbf{n} = \textbf{m}_p - \textbf{m}_d$, defined as the difference of the average magnetizations on each sublattice (denoted as $p$ and $d$ in the Lieb lattice). In the presence of a Zeeman field and global magnetization $m$, magnetic phases are further parametrized by the relative angle $\theta$ between $\textbf{n}$ and $\textbf{m}$. The phase is ferromagnetic (red) when staggered magnetization and global magnetization are parallel, $\theta = 0$, and a canted antiferromagnet (cAFM, blue) when they are orthogonal, $\theta = \pi/2$. (b) Both phases coexist at the mean-field level in the Lieb lattice for a Heisenberg model with positive coupling $J$. Ferrimagnetism is favored at a magnetization $|\textbf{m}| = 1/3$ down to the ground state, and at large temperatures $T > J$. At intermediate temperatures, the order continuously varies away from $|\textbf{m}| = 1/3$ from a ferrimagnetic to a canted AFM alignment. (c) By contrast, the square lattice shows a transition from a paramagnetic phase (PM, white) to a canted AFM phase at all $0 < |\textbf{m}| < 1$ when decreasing temperature.
    \label{fig:curie-weiss}}
\end{dfigure}

\begin{dfigure}{offset_calibration}
    \centering
    \noindent
    \includegraphics[width=3in]{"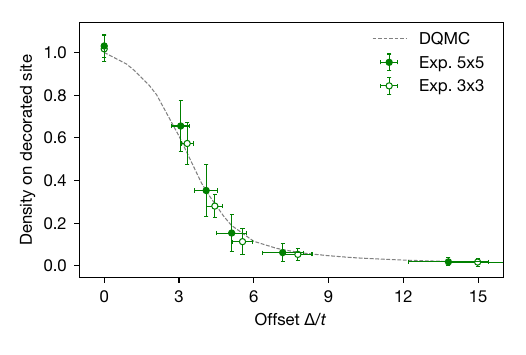"}
    \caption{\textbf{Calibration of the DMD potential.}
    Potential offset $\Delta/t$ on the decorated sites (x-axis) after fitting numerical DQMC data (dashed line) to density (y-axis) measured in the data of Fig.~4. Error bars indicate the spatial variation on the fitted offsets and decorated-site densities over the full $5\times5$ unit-cell system (closed symbols, Figs.~1 to 3) and a $3\times3$ unit-cell subset (open symbols, Fig.~4).
    \label{fig:calibrate_offset}}
\end{dfigure}

\begin{dfigure*}{disorder}
    \centering
    \noindent
    \includegraphics[width=\linewidth]{"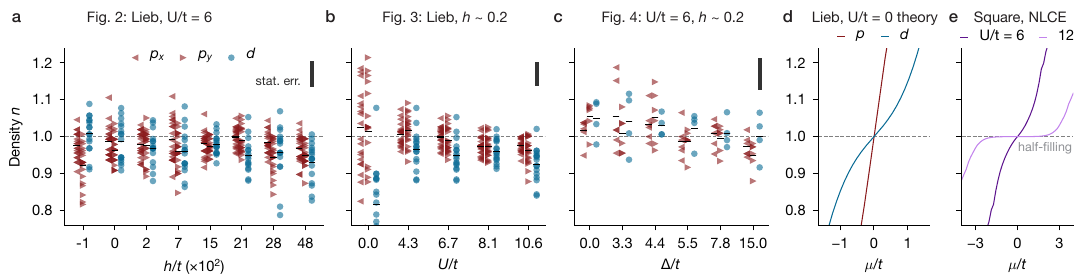"}
    \caption{\textbf{Uniformity of the Lieb lattice.} (a, b, c) Density distribution on the $p_x$-, $p_y$-, and $d$-sites for the datasets shown in Fig.~2 to~4. Each point indicates the local density on a single site, statistically averaged over a dataset, and horizontal black bars are the densities statistically and spatially averaged over a Lieb sublattice. The typical $1\sigma$ error on the single-site densities is shown at the top right corner of each panel. The averaged densities on the $p_x$, $p_y$ and $d$ sublattices are within 5\% for most datasets, with the exception of Fig.~3, in which a density of about 0.8 on the $d$-sublattice at $U/t = 0$ indicates an energy offset relative to the $p$-sublattice of order $t$. (d) Expected density $n$ on the $p$- and $d$- sublattices in the absence of interactions and at temperature $T/t = 0.4$, as a function of chemical potential $\mu/t$, zoomed around half-filling. (e) Density $n$ in the square lattice at $U/t = 6, 12$ and $T/t = 0.4$ according to numerical NLCE simulations, which is a good approximation of the average density on the Lieb lattice at large interaction strength.
    \label{fig:disorder}}
\end{dfigure*}

\begin{dfigure}{bands}
    \centering
    \noindent
    \includegraphics[width=\columnwidth]{"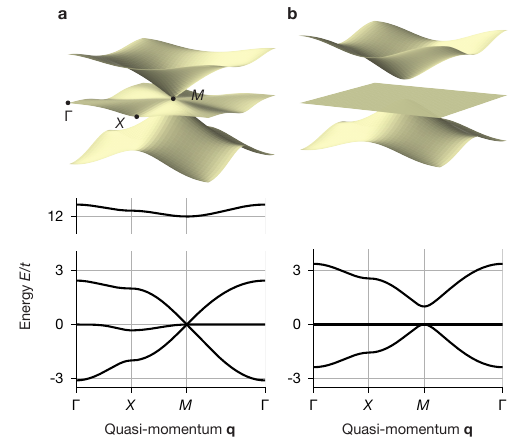"}
    \caption{\textbf{Band structures with on-site energy offsets.} (a) Introducing a fourth decorated site with finite energy offset $\Delta = 12 t$ breaks particle-hole symmetry in the three lowest bands, with a slightly dispersive second band of width $0.3 t$. (b) Adding an energy offset $\Delta_d = t$ on the $d$-sublattice preserves the flat band but opens a gap to the third band and turns the Dirac point into a quadratic band touching.
    \label{fig:band_structure}}
\end{dfigure}

\begin{dfigure}{s4}
    \centering
    \noindent
    \includegraphics[width=\columnwidth]{"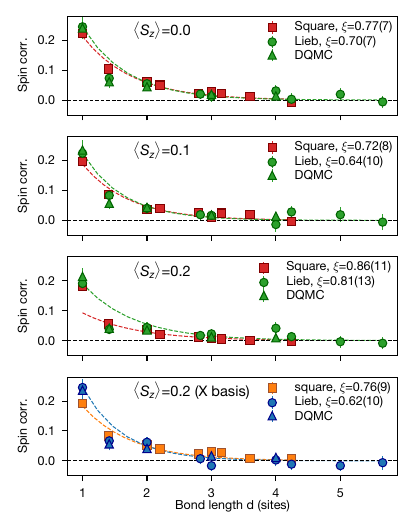"}
    \caption{\textbf{Spin correlation length.}
    Exponential fits to the spin correlations at $U/t = 6$ ($\langle S_z S_z \rangle_\text{c}$ for top three panels and $\langle S_x S_x \rangle_\text{c}$ for last panel) in the Lieb lattice and the square reservoir. For the Lieb lattice, spin correlations along four line-cuts ($d$-to-$p_x$, $d$-to-$p_y$ and the two diagonals $p_x$-to-$p_y$, similar to Ref.~\cite{costa_ferromagnetism_2016}) up to distance $4\sqrt{2}$ are used to fit the correlation length. For the square lattice, all bonds out to distance $3\sqrt{2}$ are used for fitting. 
    \label{fig:corr_length}}
\end{dfigure}

\begin{dfigure}{boundary}
    \centering
    \noindent
    \includegraphics[width=\columnwidth]{"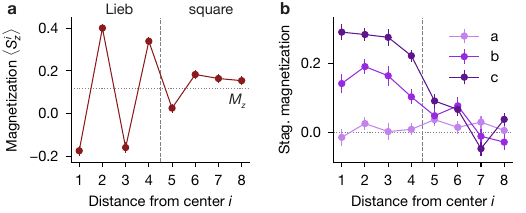"}
    \caption{\textbf{Magnetization at the Lieb-square boundary.} (a) Magnetization $\langle S_z^i \rangle$ averaged over the data of Fig.~2c across the boundary between the Lieb and the square shown in Fig.~1D. (b) The staggered magnetization $\tilde{m}_z^d$, defined in Eq.~\ref{eq:stagmag}, decays away from the boundary over a couple of sites for data of Fig.~2B, C.
    \label{fig:boundary}}
\end{dfigure}

\begin{dfigure}{ferrimagnetism_numerics}
    \centering
    \noindent
    \includegraphics[width=\columnwidth]{"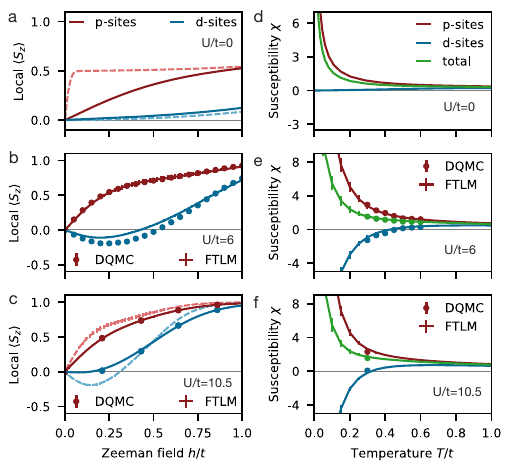"}
    \caption{\textbf{Local magnetization $\langle S_z \rangle$ and susceptibility $\chi$ at half-filling.}
    (a, b, c) Numerically computed local magnetizations for interaction strengths $U/t=0, 6,$ and $10.5$.
    For $U/t=0$, solid (dashed) lines indicate results for $T/t=0.3$ ($T/t=0.01$).
    For $U/t = 6$, the markers (solid lines) indicate results from DQMC (FTLM) simulations at $T/t=0.3$. 
    For $U/t=10.5$, the solid (dashed) lines indicate results for $T/t=0.3$ ($T/t=0.2$) from FTLM simulations while markers indicate DQMC results for $T/t=0.3$.
    (d, e, f) Numerically computed magnetic susceptibility for interaction strengths $U/t=0, 6,$ and $10.5$ as a function of temperature.
    \label{fig:ferrimagnetism_numerics}}
\end{dfigure}

\begin{dfigure}{s7}
    \centering
    \noindent
    \includegraphics[width=\columnwidth]{"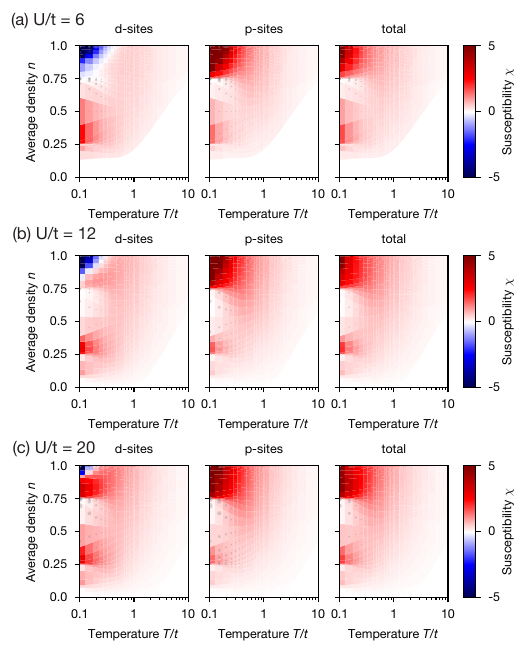"}
    \caption{\textbf{Susceptibility $\chi$ below half-filling.}
    Numerically computed susceptibility $\chi$ for $d$-sites (left), $p-$sites (middle) and their average (right) for interaction strengths (a) $U/t=6$, (b) $U/t=12$, and (c) $U/t=20$ using FTLM at average density $n$ below half-filling ($n\leq 1$). The simulations were performed for a uniform grid in chemical potential, but plotted versus density resulting in non-uniform tile sizes.
    Statistical error bars in the Lanczos method are indicated by the size of grey circles (see text).
    \label{fig:susceptibility_vs_T_n}}
\end{dfigure}

\begin{dfigure}{hartree_fock}
    \centering
    \noindent
    \includegraphics[width=\columnwidth]{"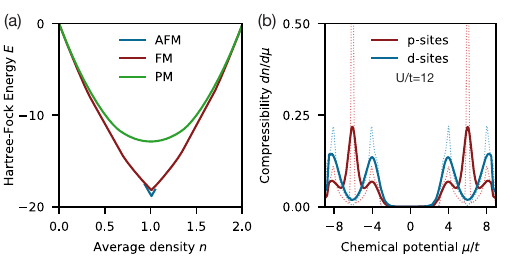"}
    \caption{\textbf{Hartree-Fock for Lieb lattice Hubbard model.}
    (a) Energies computed with Hartree Fock for antiferromagnetic ordering (AFM), ferromagnetic ordering (FM) and paramagnetic state (PM). 
    (b) Compressibility obtained from Hartree-Fock for $T/t=0.3$ (solid lines) and $T/t=0.05$ (dotted lines).
    \label{fig:hartree-fock}}
\end{dfigure}

\end{document}